\documentclass[a4paper,twocolumn,10pt,accepted=2022-05-21]{quantumarticle}
\pdfoutput=1
\usepackage[numbers,sort&compress]{natbib}

\usepackage{amsmath,amsthm,amsfonts,amssymb,mathrsfs,mathdots,graphicx,xcolor,times,xfrac,mathtools,enumerate,xr,subfigure,bbm,verbatim,comment,dsfont,soul,array,multirow}
\usepackage[T1]{fontenc}
\usepackage{comment}
\usepackage{scalerel}

\usepackage[ruled,lined]{algorithm2e}

\usepackage{color}
\definecolor{darkgreen}{RGB}{50,190,50}
\definecolor{darkblue}{RGB}{0,0,190}
\definecolor{darkred}{RGB}{238,0,0}
\definecolor{quantum}{RGB}{83,37,127}
\definecolor{quantumlight}{RGB}{169,146,191}
\definecolor{nice}{RGB}{230,0,230}
\definecolor{nicepink}{rgb}{0.858, 0.188, 0.478}
\usepackage[unicode=true,bookmarks=true,bookmarksnumbered=false,bookmarksopen=false,breaklinks=false,pdfborder={0 0 1}, backref=false,colorlinks=true]{hyperref}
\hypersetup{
	bookmarksnumbered,
	pdfstartview={FitH},
	citecolor={darkgreen},
	linkcolor={darkred},
	urlcolor={darkblue},
	pdfpagemode={UseOutlines}}
\definecolor{darkgreen}{RGB}{50,190,50}
\definecolor{darkblue}{RGB}{0,0,190}
\definecolor{darkred}{RGB}{238,0,0}
\definecolor{quantum}{RGB}{83,37,127}
\definecolor{quantumlight}{RGB}{169,146,191}
\definecolor{darkorange}{RGB}{255,100,0}
	
\usepackage{tikz}
\usetikzlibrary{shapes,backgrounds,fit,decorations.pathreplacing,arrows,decorations.markings}
\tikzstyle{vecArrow} = [thick, decoration={markings,mark=at position
   1 with {\arrow[semithick]{open triangle 60}}},
   double distance=1.4pt, shorten >= 5.5pt,
   preaction = {decorate},
   postaction = {draw,line width=1.4pt, white,shorten >= 4.5pt}]
\tikzstyle{innerWhite} = [semithick, white,line width=1.4pt, shorten >= 4.5pt]

\newcommand{\tr}{\textnormal{Tr}}

\newcommand{\ket}[1]{\ensuremath{\left|\right.\!{#1}\!\left.\right\rangle}}

\newcommand{\bra}[1]{\ensuremath{\left\langle\right.\!{#1}\!\left.\right|}}

\newcommand{\ketbra}[2]{\ensuremath{|{#1}\rangle\!\langle{#2}|}}
\newcommand{\brakket}[3]{\ensuremath{\langle{#1}|{#2}|{#3}\rangle}}

\newcommand{\proj}[1]{\ket{#1}\!\!\bra{#1}}
\newcommand{\nr}{\ensuremath{\hspace*{0.5pt}}}

\newcommand{\suptiny}[3]{\ensuremath{^{\hspace{#1 pt}\protect\raisebox{#2 pt}{\tiny{$ #3$}}}}}

\DeclareMathOperator{\diag}{diag}
\def\ii{\mathrm{i}}
\def\tr{\mathrm{Tr}}

\begin{document}

\title{Metrology-assisted entanglement distribution in noisy quantum networks}
\author{Simon Morelli}
\email{simon.morelli@oeaw.ac.at}
\affiliation{Institute for Quantum Optics and Quantum Information --- IQOQI Vienna, Austrian Academy of Sciences, Boltzmanngasse 3, 1090 Vienna, Austria}
\affiliation{Atominstitut,  Technische  Universit{\"a}t  Wien,  1020  Vienna,  Austria}
\orcid{0000-0001-6344-0239}
\author{David Sauerwein}
\email{sauerwein.david@gmail.com}\thanks{This work was done prior to joining AWS.}
\affiliation{Amazon Web Services Europe, Z{\"u}rich, Switzerland}
\author{Michalis Skotiniotis}
\email{michail.skoteiniotis@uab.cat}
\affiliation{F{\'i}sica Te{\`o}rica: Informaci{\'o} i Fen{\`o}mens Qu{\`a}ntics, Departament de F{\'i}sica, Universitat Aut{\`o}noma de Barcelona, 08193 Bellaterra, Spain}
\orcid{0000-0001-6935-7460}
\author{Nicolai Friis}
\email{nicolai.friis@univie.ac.at}
\affiliation{Institute for Quantum Optics and Quantum Information --- IQOQI Vienna, Austrian Academy of Sciences, Boltzmanngasse 3, 1090 Vienna, Austria}
\affiliation{Atominstitut,  Technische  Universit{\"a}t  Wien,  1020  Vienna,  Austria}
\orcid{0000-0003-1950-8640}

\begin{abstract}
We consider the distribution of high-dimensional entangled states to multiple parties via noisy channels and the subsequent probabilistic conversion of these states to desired target states using stochastic local operations and classical communication.
We show that such state-conversion protocols can be enhanced by embedded channel-estimation routines at no additional cost in terms of the number of copies of the distributed states. 
The defining characteristic of our strategy is the use of those copies for which the conversion was unsuccessful for the estimation of the noise, thus allowing one to counteract its detrimental effect on the successfully converted copies.
Although this idea generalizes to various more complex situations, we focus on the realistic scenario, where only finitely many copies are distributed and where the parties are not required to process multiple copies simultaneously.
In particular, we investigate the performance of so-called one-successful-branch protocols, applied sequentially to single copies and an adaptive Bayesian estimation strategy.
Finally, we compare our strategy to more general but less easily practically implementable strategies involving distillation and the use of quantum memories to process multiple copies simultaneously. 
\end{abstract}

\maketitle


\section{Introduction}\label{sec:introduction}

Entanglement between two or more parties is an important ingredient for the development of many quantum technologies such as, e.g., fault-tolerant quantum computation~\cite{Scott2004, HiroshimaHayashi2006, DuerBriegel2007, BrunHsieh2013}, quantum simulation~\cite{MartyCramerPlenio2016, FriisMartyEtal2018, DalmonteVermerschZoller2018}, and, in particular, communication via quantum networks~\cite{EppingKampermannMacchiavelloBruss2017, BaeumlAzuma2017, PivoluskaHuberMalik2018, RibeiroMurtaWehner2018,PompiliEtAl2021}. A landmark goal of quantum technologies is the establishment of a `quantum internet'~\cite{Kimble2008,WehnerElkoussHanson2018,Cacciapuoti2019}---a highly interconnected quantum network able to distribute and manipulate entangled quantum states via fibres and optical links. Recent efforts~\cite{Duer2017} go in the direction of understanding the resources, challenges and opportunities that come along with such an endeavour. In this context, entanglement is a crucial resource that allows spatially separated parties to overcome the restriction of local operations and classical communication (LOCC) thereby implementing classically impossible tasks, such as quantum teleportation~\cite{BennettEtAl1993, Bouwmeester-etal1997, Llewellyn-Thompson2020}. However, the technological requirements for quantum processors and networks on a large scale are immense and will have to involve significant improvements in the efficient manipulation and control of quantum systems. A deeper comprehension of the possibilities and limitations for multipartite entanglement distribution and its manipulation via LOCC is hence necessary. The present work aims to contribute to this endeavour by exploring the synergies between stochastic LOCC (SLOCC) protocols---that aim to convert multiparty quantum states---and quantum parameter estimation protocols whose goal is to improve entanglement distribution by precisely identifying the noise suffered by such states when distributed through quantum networks.

On the theoretical side, a lot of progress has been made with regards to the conversion of multipartite entanglement via LOCC (cf.~\cite{DuerVidalCirac2000, AcinBrussLewensteinSanpera2001, DeVicenteSpeeKraus2013, SchwaigerSauerweinCuquetDeVicenteKraus2015, DeVicenteSpeeSauerweinKraus2017, SpeeDeVicenteSauerweinKraus2017, SauerweinWallachGourKraus2018}), despite the fact that LOCC is notoriously difficult to handle mathematically~\cite{ChitambarLeungMancinskaOzolsWinter2014}, while practical developments towards efficient entanglement certification techniques (see, e.g.,~\cite{BavarescoEtAl2018, Lu-Pan2018, FriisVitaglianoMalikHuber2019, ZhouZhaoYuanMa2019, Herrera-ValenciaSrivastavPivoluskaHuberFriisMcCutcheonMalik2020, MooneyWhiteHillHollenberg2021b}) and real-world quantum networks are rapidly progressing (cf.~\cite{ValivarthiEtAl2020}). Yet, basic theoretical questions about multipartite entanglement structures arising in networks have only begun to be explored~\cite{NavascuesWolfeRossetPozasKerstjens2020, KraftDesignolleRitzBrunnerGuehneHuber2021, KraftSpeeYuGuehne2021, AbergNeryDuarteChaves2020, SpeeKraft2021}. In particular, apart from recent examples~\cite{YamasakiMorelliMiethlingerBavarescoFriisHuber2021, NevenGunnHebenstreitKraus2021}, mostly transformations of pure states in the single-copy regime or the asymptotic limit of infinitely many copies have been considered. 

Here, we consider a scenario that more closely resembles situations expected to be encountered in future real-world quantum networks: the distribution of multiple but finitely many copies of multipartite quantum states via imperfect or varying channels available only for a restricted period of time.
There, one central node, the vendor, is assumed to be capable of implementing the entangling operations necessary to create a highly entangled quantum state that is subsequently distributed to spatially separated parties via imperfect but local quantum channels. In general, the states provided by such a vendor will not match the target state that the parties might wish to employ in their communication or computation protocol of choice.
For multiple parties, there is no unique resource state that can be deterministically converted to any other state via LOCC; in fact, the so-called maximally entangled set (MES) of states from which any state can be reached in this way has infinitely many elements~\cite{DeVicenteSpeeKraus2013}.
In addition, the vendor might only be able to create a limited variety of such states and may therefore be unable to offer the desired state. Indeed, this might even be in the interest of the users, as they may not want to divulge information on the specific form (and hence the intended use) of their desired target state. Since deterministic LOCC transformations between multi-party pure states (with fixed local dimensions) almost never exist~\cite{GourKrausWallach2017, SauerweinWallachGourKraus2018}, the distributed state will have to be converted to the target state \emph{probabilistically}, i.e., via SLOCC.
Two states that can be converted into each other via SLOCC are said to belong to the same \emph{entanglement class}.

Optimal SLOCC conversion protocols will generally require acting on many copies simultaneously. Beside the fact that little is known about the structure of such protocols yet, the implementation of complex multi-copy operations also represents a substantial technological challenge.
Therefore we focus on sequential strategies that process a single copy at a time. Compared to general protocols the difficulty to implement a sequential strategy does not increase with the number of copies.
To employ  a finite number, $k_{\text{s}}$, of copies for multipartite quantum communication or computation, users must purchase, on average, a larger number $k=k_{\text{s}}/p_{\text{s}}$ of copies from the vendor to compensate for the nonzero failure probability $p_{\text{f}}=1-p_{\text{s}}>0$ of the conversion protocol.

Here, we demonstrate that this apparent disadvantage has important redeeming qualities in the presence of noise. We show that, for certain `benign' types of noise, those copies for which the state conversion was unsuccessful can nevertheless be useful for obtaining information about the noise and hence partially compensate its effects on the successfully converted copies. Our results thus establish that multipartite entanglement distribution protocols can be assisted by quantum metrology at no additional resource costs in terms of additionally distributed states, and thus feature built-in noise robustness `for free'. 
  
To reach this conclusion, we first consider a particular situation as described above, i.e., a family of specific noisy multipartite state conversion protocols and show how the use of parameter estimation tools leads to an improvement in the quality of the resulting final states. Based on this exemplary situation, we then discuss the applicability of this approach to more general entanglement distribution protocols and contrast it with alternative strategies based on distillation. Finally, we discuss the merits of metrology-assisted entanglement conversion protocols and argue for their default integration in future quantum networks.

\section{Noisy entanglement distribution}\label{sec:noisy_ent_dist}

We consider a situation where $N$ parties require several copies of a highly entangled $N$-qudit state $\rho_{\text{target}}$ for a specific task.
They purchase a number of copies from a vendor, where the purchased state $\rho_{\text{vendor}}$ differs from the target state $\rho_{\text{target}}$, but belongs to the same entanglement class and can thus be converted to the target state via SLOCC.
The vendor distributes the state $\rho_{\text{vendor}}$ to the $N$ parties via local quantum channels represented by completely-positive, trace-preserving (CPTP) maps, 
$\{\Lambda_i\}_{i=1}^N$, so that the state shared by the $N$ parties is given by
\begin{equation}
\rho_{\text{received}}=\Lambda(\rho_{\text{vendor}})=\bigotimes_{i=1}^N\Lambda_i(\rho_{\text{vendor}}).
\label{eq:LO_channel}
\end{equation}
The parties now wish to transform the received state $\rho_{\text{received}}$ into the target state $\rho_{\text{target}}$ within the same entanglement class. This can be done either deterministically or probabilistically.
We focus on transformations of fully entangled (i.e., with full local rank) pure states via a so-called one-successful-branch protocol (OSBP)~\cite{AcinJaneDurVidal2000,SauerweinSchwaigerKraus2018}. The OSBP is in fact the optimal SLOCC protocol for almost all $N$-qudit pure-state transformations~\cite{SauerweinSchwaigerKraus2018}. However, the idea of using unsuccessful branches of SLOCC protocols for parameter estimation presented in this paper applies in equal measure to any other probabilistic transformations of pure and mixed states of arbitrary dimensions.

In an OSBP that transforms $\ket{\Psi}$ into $\ket{\Phi}$, each party $k$ performs
a single two-outcome measurement $\{M_{\text{s}}^k,\, M_{\text{f}}^k\}$ with $\sum_{i=\text{s,f}}(M_{i}^k)^{\dagger}M_{i}^k=\mathds{1}$. The successful outcome $s$ leads to a state that can still be transformed to the final state, while the failure outcome $f$ leads to a state that is no longer fully entangled and can therefore no longer be transformed into $\ket{\Phi}$. Hence, the final state is only reached if the successful measurement outcome is realised for all parties. That is, only one branch of the OSBP is successful, as suggested by its name, and realized with some probability $p_{\text{s}}$.

The transformation performed by the $N$ parties to convert the received state into the target state depends on the channel used to distribute the state.
For the general case when the channel is not sufficiently known, the task for the parties becomes more challenging.
They have to estimate the channel, i.e., identify the channel $\Lambda$ from a discrete or continuous family of possible candidates, before transforming the received state $\rho_{\text{received}}$.
The goal of the $N$ parties is now to obtain the largest number of copies $\rho_{\text{corr}}$ that are $\epsilon$-close in trace distance to the desired target state $\rho_{\text{target}}$, i.e.,
\begin{align}
    D(\rho_{\text{corr}},\rho_{\text{target}})=\frac{1}{2}\tr\Bigl[\sqrt{(\rho_{\text{corr}}-\rho_{\text{target}})^2}\,\Bigr]\,\,\le\,\epsilon.
\end{align}

If an asymptotically large number of copies of the state is available, the parties can always sacrifice a finite number of such states to estimate the channel to any desired precision before commencing with the state transformation. Consequently, the incurred cost for estimation has no bearing on the final yield of the target state. On the other hand, if only a finite number of copies is available, then the above strategy can be quite wasteful, and protocols that estimate the channel whilst implementing the OSBP at the same time may provide better yields. This is due to the observation that the states resulting from the failure branches of the OSBP potentially carry information about the noisy channel and can thus be used to estimate the latter at no extra cost.  

A particular class of channels for which simultaneous channel estimation and state conversion may be beneficial is the family of channels that admit a decomposition into Kraus operators that commute with the POVM corresponding to the OSBP.
Such channels include dephasing and phase drift~\cite{Jarzyna2014} noise which are typical types of noise in quantum communication using optical fibres~\cite{Wanser1992}.  Alternatively, such channels also describe a situation often encountered in satellite communication where the sender (vendor) and receivers ($N$ parties) do not share a common phase reference during the transmission stage~\cite{Fanizza2020}. In this latter case the receivers know that the state they receive is local-unitarily (LU) equivalent to the original state of the vendor and thus is in the same entanglement class.  However, the lack of knowledge of the relative phase between the sender's and receivers' phase reference makes channel estimation indispensable, especially if the task they have in mind requires a particular state $\rho_{\text{target}}$, and not any other local-unitarily equivalent state.

\begin{figure}[t]
      \includegraphics[width=0.95\columnwidth]{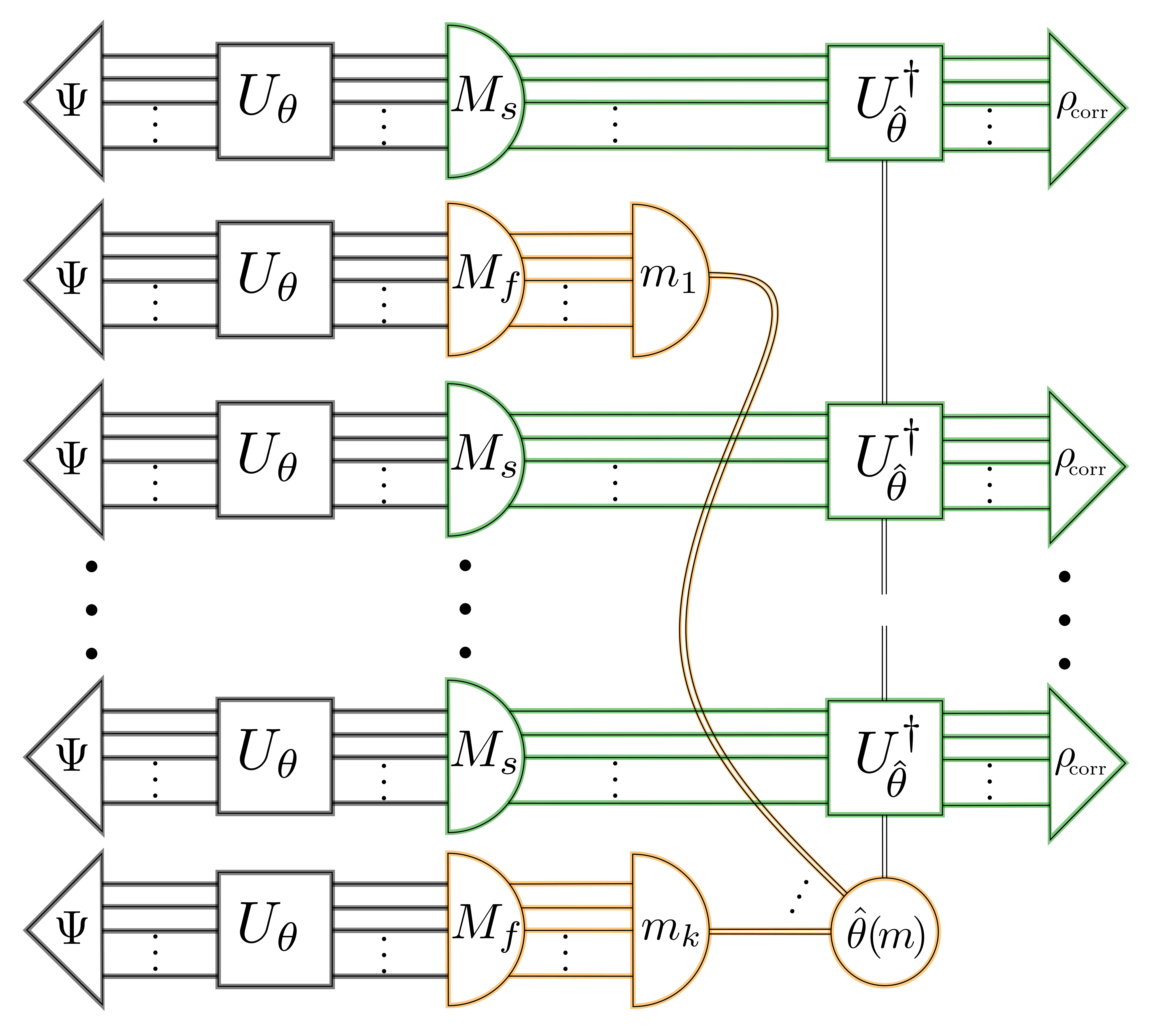}
        \caption{
        Schematic of the metrology-assisted entanglement distribution protocol. 
        Each copy of the distributed multi-party state $\ket{\Psi}$ is subject to a (local) noise channel (here represented by an unknown unitary $U_{\theta}$) parameterized by $\theta$, which is subsequently measured by an OSBP with measurement operators $M_{\text{s}}$ and $M_{\text{f}}$, corresponding to successful (green) and failed (orange) state conversion, respectively. The copies in the failure branch can be used to perform further measurements with outcomes $m_{i}$ to estimate the noise channel, and apply corresponding corrections (represented here by $U_{\hat{\theta}}^{\dagger}$) to obtain partially corrected states $\rho_{\text{corr}}$.
        }
    \label{fig:circuit}
\end{figure}

In what follows, we shall consider the above class of channels as a special case of more general parameter estimation problems and employ a Bayesian estimation strategy to estimate the parameter (or set of parameters) $\theta$ that characterizes the channel, illustrated in Fig.~\ref{fig:circuit}.
Such a strategy involves one or more rounds of subsequent measurements on different copies of the received states.
The knowledge of the parameter, $\theta$, is updated in every round depending on the measurement outcome, $m$, according to Bayes' law, i.e.,
\begin{align}
        p(\theta|m) &=\,\frac{p(m|\nr\theta)\,p(\theta)}{p(m)}.\label{bayeslaw}
    \end{align}
Based on the posterior distribution of $\theta$ one can then assign an estimator $\hat{\theta}(m)$ after each round and adapt the next measurement to this estimate.
This makes Bayesian parameter estimation more flexible and better suited for a limited number of probes compared to frequentist estimation.
For an increasing number of copies, also a frequentist strategy can be employed to estimate the channel from the failure branches of the OSBP. 
Here, however, we will focus on scenarios with a very restricted number of copies. 
In particular, we consider the exemplary situation of local dephasing noise, 
to provide quantitative support for our strategy of metrology-assisted entanglement distribution.

\section{Entanglement distribution with dephasing noise}\label{sec:dephasing_noise}

Assume the state distributed by the vendor is a pure $N$-qudit state from the GHZ class of the form
\begin{equation}
\ket{\Psi}=\sum_{l=0}^{d-1}\sqrt{\lambda_l}\ket{l}^{\otimes N},
\label{eq:vendor_state}
\end{equation}
where we assume without loss of generality that 
$\lambda_m \ge \lambda_{m+1}$.
Let us further assume that the vendor distributes these states via $N$ local dephasing channels, i.e., each qudit undergoes a unitary transformation parameterized by an unknown angle $\theta$,
\begin{equation}
U(\theta)=\sum_{l=0}^{d-1}e^{\ii l\theta}\proj{l}.
\label{eq:unitaries}
\end{equation}
The parties wish to obtain the maximally-entangled qudit state 
\begin{equation}
\ket{\mathrm{GHZ}_d}=\sqrt{\tfrac{1}{d}}\sum_{l=0}^{d-1}\ket{l}^{\otimes N}.
\label{eq:qutrit_GHZ}
\end{equation}
To this end, they estimate the unknown phase $\theta$ of the dephasing noise acting on the received states and recover the target state via SLOCC-transformations from the received state.

More specifically, let us assume the vendor produces the pure two-parameter qutrit state 
\begin{align}
\ket{\Psi(\alpha,\beta)}=&\sin\alpha\cos\beta\ket{0}^{\otimes N}+\sin\alpha\sin\beta\ket{1}^{\otimes N}
\label{eq:vendor_state2}\\
&+\cos\alpha\ket{2}^{\otimes N},\notag
\end{align}
where $\tan^{-1}(\csc\beta)\le\alpha\le\pi/2$ and $0\le\beta\le\pi/4$. This choice of angles ensures that 
$\sin\alpha\cos\beta\geq\sin\alpha\sin\beta\geq\cos\alpha\ge0$.  The parties wish to obtain the maximally entangled state of 
Eq.~\eqref{eq:qutrit_GHZ}, for which the optimal OSBP comprises measurement operators 
\begin{align}
M_{\mathrm{s}}&=\diag(\cot\alpha\sec\beta,\cot\alpha\csc\beta,1),\label{eq:OSBP}\\
M_{\mathrm{f}}&=\diag(\sqrt{1-\cot^2\alpha\sec^2\beta},\sqrt{1-\cot^2\alpha\csc^2\beta},0)
\nonumber
\end{align}
for one party and the identity for the remaining parties.
The conversion is successful with probability $p_{\text{s}}=3\cos^2\alpha$,
which is already the optimal conversion rate for any protocol acting on a single copy of the state.
This follows from the fact that the optimal conversion rate from a bipartite pure state $\ket{\Psi}$ to another bipartite pure state $\ket{\tilde{\Psi}}$ can be calculated~\cite{Vidal1999} from the respective Schmidt coefficients $\lambda_{i}$ and $\tilde{\lambda}_{j}$ (real, non-negative,and in decreasing order) as $P(\Psi\rightarrow\tilde{\Psi})=\min_{j}\sum_{l=j}\lambda^{2}_{l}/\sum_{l'=j}\tilde{\lambda}^{2}_{l'}$, and from the observation that the optimal conversion rate for any splitting of the $N$ parties into two groups gives an upper bound on the conversion rate. By splitting the state in Eq.~(\ref{eq:vendor_state2}) into one party versus the remaining $N-1$ parties, one immediately sees that the conversion rate for a single state is bounded by $P(\Psi(\alpha,\beta)\rightarrow\mathrm{GHZ}_{3})=3\cos^2\alpha$ for the parameter range we have chosen for $\alpha$ and $\beta$.

With probability $p_{\text{f}}=1-3\cos^2\alpha$ the conversion protocol 
fails and results in the state 
\begin{equation}
\ket{\Phi_{\text{f}}}=a\ket{0}^{\otimes N}+\sqrt{1-a^2}\ket{1}^{\otimes N},
\label{eq:fail_state}
\end{equation}
where 
\begin{equation}
a=\frac{\sqrt{\sin^2\alpha\cos^2\beta-\cos^2\alpha}}{\sqrt{1-3\cos^2\alpha}}.
\end{equation}
Observing that
$[M_{x},U(\theta)]=0$, $\forall\,x\in\{\text{s, f}\}$,
it follows that the resulting states for the successful and failure branches of the OSBP are given by 
\begin{subequations}
\begin{align}
\ket{\mathrm{GHZ}_3(\theta)}&=\sqrt{\frac{1}{3}}\sum_{l=0}^2e^{\ii lN\theta}\ket{l}^{\otimes N},\\
\ket{\Phi_{\text{f}}(\theta)}&=a\ket{0}^{\otimes N}+\sqrt{1-a^2}e^{\ii N\theta}\ket{1}^{\otimes N},
\label{eq:noisy_states}
\end{align}
\end{subequations}
respectively, with the corresponding probabilities invariant under the presence of dephasing noise.

In the case that the OSBP is successful, the parties obtain the state
\begin{align}\nonumber
\Lambda\bigl[\rho_{\mathrm{GHZ}_3}\bigr]&=\int_0^{2\pi}p(\theta)U(\theta)^{\otimes N}
\rho_{\mathrm{GHZ}_3}
U(\theta)^{\dagger\,\otimes N}\mathrm{d}\theta,
\label{eq:scenario_2}
\end{align}
where $\rho_{\mathrm{GHZ}_3}=\proj{\mathrm{GHZ}_3}$ and $p(\theta)$ is a probability distribution over $\theta\in[0,2\pi)$ describing the belief/knowledge the parties have of\footnote{Actually the estimation only gives us information about $N\theta$, which allows us to draw conclusions about $\theta$ only up to a shift by $2\pi/N$. But this is not a problem, since the final state also only depends on $N\theta$.} $\theta$. Loosely speaking, the more peaked the distribution becomes, the closer the state gets to the target state. 
The goal of the $N$-parties now is to apply correction operations to the successfully converted copies to obtain the highest yield of $N$-partite highly entangled qutrit states $\rho_{\text{corr}}$, that are $\epsilon$-close in trace distance to the ideal state $\ket{\mathrm{GHZ}_3}$.

In the case of a flat prior, $p(\theta)=(2\pi)^{-1}$, we can analytically calculate (see Appendix~\ref{app: single measurement} for details) the average trace distance to the target state after the first measurement to be
\begin{small}
\begin{align}
    \bar{D}(\rho_{\text{corr}},\rho_{\text{target}})
    =\tfrac{1}{6}\bigl(1+\sqrt{9+8a^2(1-a^2)-16a\sqrt{1-a^2}}\bigr).
\end{align}
\end{small}
The average distance after one measurement hence only depends on the parameter $a$. Since  $a\in[0,1]$, the maximal value $\tfrac{2}{3}$ is obtained for $a=0$ or $a=1$. This coincides with the distance based on the prior, i.e., ignoring measurement data. The minimal value of $\bar{D}=\frac{1}{6}(1+\sqrt{3})$ is obtained for $a=1/\sqrt{2}$.

Given a particular measurement outcome the posterior distribution 
is updated using Bayes' law, [Eq.~(\ref{bayeslaw})].
In general the parties can optimize their measurement directions at every step but this amount of generality quickly renders the problem analytically intractable. We thus resort to numerical analysis henceforth, and we focus on a strategy where the measurement direction on the Bloch sphere of the two-dimensional subspace relevant for the copies in the failure branch (see Appendix~\ref{app: single measurement}) is chosen on the equator but perpendicular to the estimator of $\theta$ of the previous round. This measurement is particularly sensitive in the direction of $\hat{\theta}$, since both outcomes are equally likely in this case. The simplest strategy of fixed local measurements is treated in Appendix~\ref{app: multiple copies}, where also strategies that change the measurement direction according to some pre-established rule are considered.

\begin{figure}[t]
     \includegraphics[width=0.95\columnwidth]{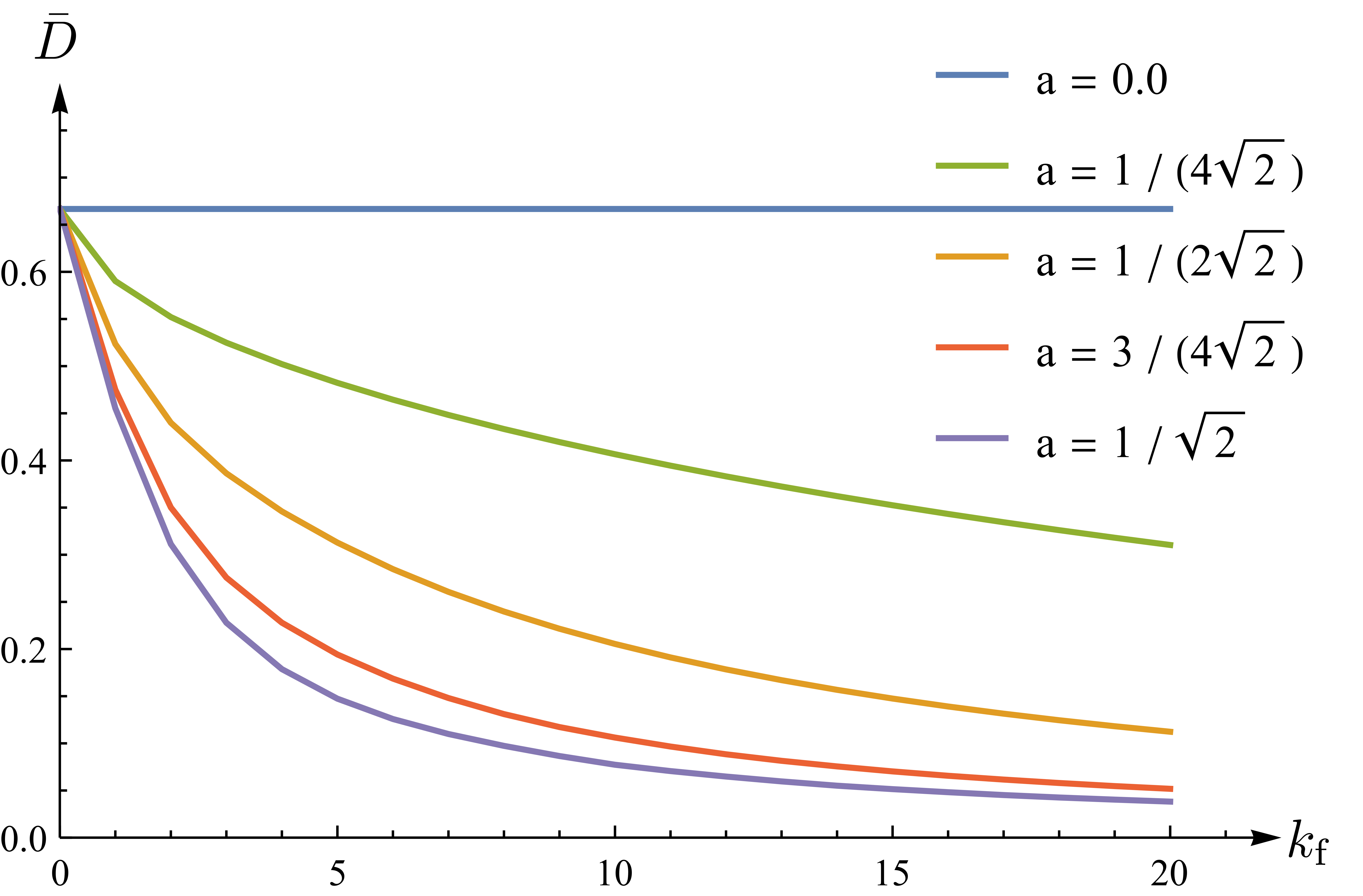}
        \caption{
        Average distance to the target state. 
        The curves show the average trace distance (vertical axis) between the target state and the successfully converted copies for the exemplary metrology-assisted entanglement distribution protocol discussed in Sec.~\ref{sec:dephasing_noise} for different values of $a$ as a function of the number $k_{\text{f}}$ of copies (horizontal axis) in the failure branch.
        }
    \label{fig:distance2}
\end{figure}

In Fig.~\ref{fig:distance2} we see how the trace distance $\bar{D}$ of the corrected state resulting from the successful branch to the target state reduces with the number of failure branches used for the estimation. After 20 measurement rounds the average trace distance drops to $\bar{D}\approx0.038$ for $a=1/\sqrt{2}$. This is a remarkable improvement considering that the failure states would be discarded in the naive protocol that we will discuss in Sec.~\ref{sec:comparison}.

One could also think of more sophisticated strategies, where a swap operation is applied to the state by the vendor before it is distributed, changing the basis vectors $\ket{1}\leftrightarrow\ket{2}$. Since the state $\ket{2}$ reacts more sensitive to the transformation, i.e., picks up a phase of $e^{2\ii N\theta}$, this can be advantageous for estimation. Note, however, that in this case one can draw conclusions about $\theta$ only up to a $\pi/N$ phase shift, whereas the successful branch still depends on $N\theta$. Thus, this strategy can only be employed once the interval of $\theta$ is sufficiently narrowed down.

Although we focus on the distribution of qutrit GHZ states under local dephasing noise in our example, we want to emphasize that this situation can be straightforwardly generalized to the distribution of any $d$-dimensional state in the GHZ entanglement class of the form of Eq.~(\ref{eq:vendor_state}) with the same noise model. Here an OSBP converting the received state to the GHZ state in Eq.~(\ref{eq:qutrit_GHZ}) leaves the failure-branch in a subspace of dimension  $2\le d_{\text{f}}<d$ and one can use estimation strategies similar to that used in Appendix~\ref{app: good copy estimation}.
Additionally, the GHZ class is broad in the sense that, according to Ref.~\cite{DuerVidalCirac2000}, almost all pure multipartite entangled three-qubit states belong to this entanglement class, whereas states from the W class form a subset of measure zero. While this may not necessarily hold for higher dimensions or more parties, our example remains relevant as states in the GHZ class are often used (see, e.g., \cite{WangEtAl2018,Gong-Pan2019a,MooneyWhiteHillHollenberg2021,PogorelovEtAl2021}) and have desirable properties such as their symmetry with respect to the exchange of parties.

\section{Comparison with 'naive' protocols}\label{sec:comparison}

In this section we compare the performance of our strategy to what we call 'naive' strategies. These are strategies that first sacrifice a number of copies of the received state for the estimation of the LU-channel and subsequently perform the transformation to the target state for all following copies.
For this comparison we are interested in determining the number $k$ of copies of received states required to obtain a certain number $k_{\text{s}}$ of successfully converted states whose average trace distance from the desired target state is below a given value $\epsilon$. Moreover, to make the comparison fair, first note that our previous metrology-assisted entanglement conversion protocol processes the received multipartite systems \emph{sequentially}, i.e., one copy of each multipartite state at a time, and only uses local measurements and operations both for the estimation and the state transformation. Although this drastically simplifies potential practical implementations, it is known that processing multiple copies simultaneously \textemdash\ via the use of quantum memories \textemdash\ can lead to a higher success rate for entanglement distillation~\cite{DattaLeditzky2015, FangWangTomamichelDuan2019}. For a fair comparison we therefore impose the same restrictions on the 'naive' strategy, i.e. we assume that also any 'naive' protocol is restricted to process each copy of the multipartite state sequentially via LOCC.

Within this framework for the 'naive' approach we consider a Bayesian strategy that estimates the phase parameter based on measurements performed on a certain number $k_{\text{e}}$ of individual received states, while the $k-k_{\text{e}}$ subsequent copies are individually transformed via the previously considered SLOCC protocol. We calculate the average trace distance of the successfully converted copies from the target state as a function of the initially sacrificed $k_{\text{e}}$ copies used for estimation. Since the chosen Bayesian strategy might not be optimal in general, we supplement this analysis with an estimate of the lower bound on the achievable trace distance based on the Bayesian quantum Cram\'er-Rao bound. Comparing these results to the performance of our previously presented metrology-assisted protocol, we conclude that there exist combinations of the parameters $k$ and $\epsilon$ for which our protocol outperforms any ‘naive’ protocol since the latter always sacrifices some of the copies that can be successfully converted within an approach that solely uses the failure-branch states for estimation.

After these 'fair' comparisons, we also relax the condition of sequential processing and consider protocols based on distillation where each party can use quantum memories and is able to process multiple copies at the same time. For this scenario, we derive upper bounds on the obtainable number of successfully converted copies valid for any distillation strategy and conclude that such strategies potentially outperform our sequential strategy. As these bounds are only asymptotically tight, however, it remains unclear how well the optimal strategy actually performs on a finite number of copies. 

Let us now consider the task of the previous example, i.e., $k$ copies of the state in Eq.~(\ref{eq:vendor_state2}) are distributed via a local dephasing channel, described in Eq.~(\ref{eq:unitaries}), and the goal of the parties is to obtain the highest number of states (at least) $\epsilon$-close to $\ket{\mathrm{GHZ}_3}$. They therefore use $k_{\text{e}}$ copies of $\ket{\Psi(\alpha,\beta)}$ to estimate the phase $\theta$ and assign an estimator $\hat{\theta}(\mathbf{m})$ based on the measurement outcome $\mathbf{m}=(m_1,\dots,m_{k_\text{e}})$.
Using a Bayesian estimation strategy as before, we have to choose an appropriate measurement. As our probe is a three-level system now, we perform a three-outcome projective measurement. Since we assume not to have any knowledge of $\theta$ initially, we consider a flat prior, for which a reasonable choice of measurement basis presents itself in the form of any basis that is mutually unbiased with respect to the eigenbasis of the generating Hamiltonian of the local dephasing noise $U(\theta)$.
Since no further restrictions seem to apply to the measurement basis we therefore use the discrete Fourier transform of the eigenbasis of the generator~\cite{BrierleyWeigertBengtsson10}.
However, note that for the measurement on the second and subsequent copies, we rotate the measurement basis by $U(\hat{\theta}+\pi/3)$ to take into account our estimate $\hat{\theta}$.
The obtained basis still is mutually unbiased with respect to the eigenbasis of the encoding Hamiltonian. A more detailed calculation of this estimation can be found in Appendix~\ref{app: good copy estimation}.

Let us compare this strategy to the numerical results of the previously employed estimation of the failure branch for the exemplary parameters $k=100$, $\alpha=\arccos(2/\sqrt{15})$ and $\beta=\pi/4$.
This leads to $p_{\text{s}}=4/5$, and so to $k_{\text{s}}=80$ and $k_{\text{f}}=20$ successful and failure branches, respectively, on average. Twenty rounds of estimation with the failure branch lead to an average trace distance of $\bar{D}\approx0.038$, see Fig.~\ref{fig:distance2}.
To achieve a comparable distance with the received state requires $k_{\text{e}}=9$ copies on average, see Fig.~\ref{fig:distance4} in Appendix~\ref{app: good copy estimation}. 
The observation that fewer copies of the received state are needed to estimate the phase to the same precision as compared to the failure-branch state comes as no surprise, 
since any processing of the state after the transmission can only reduce (on average) the amount of information about $\theta$ encoded in the system. This means that the received states will generally be more advantageous for the estimation than states in the failure branch. While this is true, it is essentially the same strategy that was also used for the failure branch in the previous protocol adapted to the different probe state. Hence we argue that this is indeed a fair comparison, as we do not know the optimal strategy for finitely many copies.

This notwithstanding, we can calculate a lower bound on the number of copies needed for any estimation strategy to reach a given trace distance based on the Bayesian quantum Cram\'er-Rao bound, see Appendix~\ref{app: lower bound good copy estimation} for a detailed calculation.
Comparing this to the numerical results of the previously employed strategy for the same exemplary parameters $k=100$, $\alpha=\arccos(2/\sqrt{15})$ and $\beta=\pi/4$, one finds that achieving a comparable trace distance, any estimation needs on average at least $k_{\text{e}}\ge7.5$ copies of the received state, see Fig.~\ref{fig:distance4} in Appendix~\ref{app: good copy estimation}.
However, note that $k_{\text{e}}\ge7.5$ is only a lower bound, and it is not clear if a strategy exists that saturates this bound for finitely many copies.

Once the phase $\theta$ has been estimated sufficiently well, the states resulting from the successful branches can be transformed via SLOCC.
We have already argued in the previous section that any sequential strategy transforming individual copies of the state in Eq.~(\ref{eq:vendor_state2}) to the state $\ket{\text{GHZ}_3}$ is bounded by $p_{\text{s}}\le3\cos^2\alpha$.
For the same exemplary parameters considered before, this leads to a success rate of $p_{\text{s}}=4/5$. For the 'naive' Bayesian strategy we hence obtain $(k-k_{\text{e}})p_{\text{s}}=91\,\tfrac{4}{5}=72.8$ successfully converted copies on average from the initial $100$ copies of the distributed state.
Additionally, the lower bound based on the Bayesian quantum Cram\'er-Rao bound informs us that any 'naive' sequential strategy obtains at most $(k-k_{\text{e}})p_{\text{s}}=92.5\,\tfrac{4}{5}=74$ copies of the target state on average, proving our failure-branch estimation considerably superior with $80$ successfully converted copies on average in this example.

Indeed, for any noise particular SLOCC conversion protocol with success probability $p_{\text{s}}$ (in which a channel is estimated whose Kraus operators commute with the SLOCC protocol) one may find pairs of $k$ and $\epsilon$ for which our strategy outperforms naive protocols since the latter always sacrifice some of the copies that can otherwise be successfully converted. 
If the average trace distance $\bar{D}$ after $k$ copies is equal to or smaller than the desired distance $\epsilon$, then our protocol cannot be beaten by a naive strategy for this combination of parameters since any naive strategy would have to sacrifice some number $k_{\text{e}}>0$ of the received states, thus ultimately being left with only $p_{\text{s}}(k-k_{\text{e}}) < k_{\text{s}} = p_{\text{s}}k$ copies within the desired distance $\epsilon$. However, note that this does not imply that our strategy outperforms naive protocols for all values of $k$ and $\epsilon$. 
There also exist pairs of $k$ and $\epsilon$ where our failure branch estimation is not sufficiently precise and good copies need to be sacrificed for estimation to reach the desired precision. 
In such a case the performance of such a 'hybrid' strategy compared to a 'naive' strategy depends on the specific parameters of the considered example.

In addition, we note that naive protocols based on distillation, i.e., which act jointly on all $k-k_{\text{e}}$ states available after the initial estimation could in principle perform better than sequential protocols based on failure-branch estimation only, even in cases where the latter outperform sequential naive protocols. 
For pure bipartite states the asymptotic distillation rate from $\ket{\psi}$ to $\ket{\phi}$ is bounded by the entropy ratio of the reduced states for any party $i$, i.e.,
\begin{align}
    \mathcal{D}(\ket{\psi}\rightarrow\ket{\phi})=\frac{S(\ketbra{\psi}{\psi}_i)}{S(\ketbra{\phi}{\phi}_i)},
\end{align}
and this bound is tight~\cite{BennettBernsteinPopescuSchumacher1996}. For multipartite states this immediately gives the, generally not tight, asymptotic bound  
\begin{align}
    \mathcal{D}(\ket{\psi}\rightarrow\ket{\phi})=\min\limits_i\Bigl(\frac{S(\ketbra{\psi}{\psi}_i)}{S(\ketbra{\phi}{\phi}_i)}\Bigr).
\end{align}
In our case, however, the latter bound is tight asymptotically~\cite{SmolinVerstraeteWinter2005, StreltsovMeignantEisert2020} and we can calculate
\begin{align}
    \mathcal{D}(\ket{\Psi(\alpha,\beta)}\rightarrow&\ket{\text{GHZ}})=-\cos^2\alpha\log_3(\cos^2\alpha)\\
    &-\sin^2\alpha\cos^2\beta\log_3(\sin^2\alpha\cos^2\beta)\notag\\
    &-\sin^2\alpha\sin^2\beta\log_3(\sin^2\alpha\sin^2\beta)\notag.
\end{align}
For our
choice of parameters this means there might exist a distillation protocol acting on all available copies at once able to successfully distill (on average) $91.6$ copies of the desired GHZ-state out of the $k-k_{\text{e}}=92.5$ remaining copies after an optimal estimation. This clearly exceeds the $80$ successful states obtained by our adaptive strategy.
Here we again want to emphasise that these $91.6$ copies are only an upper bound. Both an optimal estimation and distillation strategy are only known to exist asymptotically and it is not clear that any strategy can attain this bound for finitely many copies.
Indeed, even if such a strategy exists, it would generally require acting on all remaining copies simultaneously.
Even if the presented metrology-assisted protocol does not saturate these 'optimal' bounds, it nonetheless shows the potential of practically easily implementable adaptive strategies acting on single copies.
In particular, the task of designing and implementing optimal strategies for estimation and especially distillation becomes extremely challenging for larger numbers of copies and may be out of reach for experimental realizations in the medium term.
Therefore it is essential to find feasible strategies that perform well and whose implementation remains tractable with growing number of copies.

\section{Metrology-assisted protocols beyond dephasing noise}\label{sec:beyond_dephasing}

A key assumption in our previous example was that the parties know that the measurement operators of the OSBP commute with the Kraus operators of the noisy channel.
If this does not hold, as is the case for depolarizing, channels with high loss, or for dephasing channels with unknown rotation axis, the order in which the POVM measurements of the OSBP and those for the estimation protocol are carried out matter.
If the channel is not known to commute with the POVM operators of the OSBP, this impacts the performance of the presented strategy. Nonetheless one could design strategies relying on the failure branch for the channel estimation.
We now present a protocol that integrates the estimation of general unknown unitary transformations $\text{SU}(d)$ into an OSBP. Possible strategies to estimate a general unitary transformation are given by~\cite{ChiribellaDarianoPerinottiSacchi2004, Kahn2007}.
We focus on how an estimation strategy can incorporate the failure branch of an OSBP for a potentially more efficient estimation, without restricting ourselves to a specific strategy.

The naive way to perform this task would again be to first use a number of copies to estimate the unknown unitary sufficiently well, such that the received state can be corrected up to within trace distance  $\bar{D}\le\epsilon$ of the target. After that, some distillation procedure is applied to recover the desired state.

\begin{algorithm}
	\caption{Metrology-integrated OSBP}
	\SetKwInOut{Input}{input}
        \SetKwInOut{Output}{output}
        \Input{$k$ copies of the state $\ket{\psi}$}
        \Output{$p_{(k)}(U)$, $\hat{U}_{(k)}$ and all saved copies $|\tilde{\psi}_s^{(i)}\rangle$}
	\Begin{
	start with uniform distribution $p_{(0)}(U)$ over $SU(d)$\\
	set estimator $\hat{U}_{(0)}=\mathds{1}$\\
	$i=1$\\
	\While{$i\le k$, i.e., there are copies left}{
	    send copy number $i$ of the state $|\psi\rangle$\\
	    obtain $U^{\otimes N}|\psi\rangle$\\
	    correct to $|\tilde{\psi}^{(i)}\rangle=(\hat{U}^\dagger_{(i-1)} U)^{\otimes N}|\psi\rangle$\\
	    construct the POVM for the OSBP: \\
	    \begin{itemize}
	        \item define $M_{\mathrm{s}}^V$ as the operator s.t. $M_{\mathrm{s}}^V V^{\otimes N}\ket{\psi}=\ket{\mathrm{GHZ}_d}$\\
	        \item $M_{\mathrm{s}}^{(i)}=\int\limits_{SU(d)}p_{(i-1)}(V)M_{\mathrm{s}}^{\hat{U}^\dagger_{(i-1)}V}
	        \mathrm{d}V$\\
	        \item $M_{\mathrm{f}}^{(i)}$ s.t. $(M_{\mathrm{s}}^{(i)})^\dagger M_{\mathrm{s}}^{(i)}+(M_{\mathrm{f}}^{(i)})^\dagger M_{\mathrm{f}}^{(i)}=\mathds{1}$\\
	    \end{itemize}
	    perform OSBP with $\{M_{\mathrm{s}}^{(i)}, M_{\mathrm{f}}^{(i)}\}$\\
	    \eIf{success}
	        {get $|\tilde{\psi}_s^{(i)}\rangle=M_{\mathrm{s}}^{(i)}|\tilde{\psi}^{(i)}\rangle$\\
	        \eIf{$\scaleto{\int\limits_{\scaleto{SU(d)}{5pt}}p_{\scaleto{(i-1)}{5pt}}(U)|\brakket{\mathrm{GHZ}_d}{M_{\mathrm{s}}^{(i)}(\hat U^\dagger_{\scaleto{(i-1)}{5pt}} U)^{\otimes N}}{\psi}|^2 \mathrm{d}U
	        \ge 1-\epsilon}{15pt}$}
	            {keep $|\tilde{\psi}_s^{(i)}\rangle$ and save corresponding $M_{\mathrm{s}}^{(i)}$ and $\hat{U}_{(i-1)}$\\
	            leave probability distribution and estimator of the unitary unchanged\\
	            $p_{(i)}(U)=p_{(i-1)}(U)$\\
	            $\hat{U}_{(i)}=\hat{U}_{(i-1)}$}
	            {perform a measurement on $|\tilde{\psi}_s^{(i)}\rangle$ to estimate $U$\\
	            update distribution $p_{(i)}(U)$ and estimator $\hat{U}_{(i)}$}}
	       {perform a measurement on $|\tilde{\psi}_f^{(i)}\rangle=M_{\mathrm{f}}^{(i)}|\tilde{\psi}^{(i)}\rangle$ to estimate $U$\\
	       update distribution $p_{(i)}(U)$ and estimator $\hat{U}_{(i)}$}
	       $i++$
	    }
        \Return $p_{(k)}(U)$, $\hat{U}_{(k)}$ and all saved copies $|\tilde{\psi}_s^{(i)}\rangle$
    }
    \label{alg1}
\end{algorithm}

In our approach we combine both tasks into one. The idea is to construct the POVM operators for the OSBP based on the knowledge of the unknown unitary at that specific stage of the protocol.
We start with the POVM $\{\mathds{1},0\}$, that will always result in the success branch. Then we check if the state is sufficiently close to the target state, based on our knowledge of the unitary. This will not be the case if the distribution $p(U)$ is sufficiently broad.
If the state is not close enough to the target, we make a measurement and update our believe and the estimator.
With increasing knowledge of the unitary transformation, also the POVM of the OSBP will change and converge towards the optimal POVM for the received state.
During this process the copies in the successful branch close to the target (according to the knowledge at the time) are stored, while all the others are used for estimation. With an increasingly peaked distribution $p(U)$ also the ratio of stored copies to discarded copies (used for estimation) will increase.
Given that enough copies were purchased, we can assume that the unitary transformation is known sufficiently well at the end of this protocol.
For all stored copies it is checked if they are indeed close enough to the target according to the updated knowledge of $U$. If not, they are transformed into the target state via SLOCC.
The exact estimation and updating procedure are left unspecified here on purpose because these will strongly depend on the exact state and noise model.
In Algorithm~\ref{alg1} we consider the exemplary situation where the desired state is an $N$-partite GHZ-state of local dimension $d$ and the distributed state is of the form of Eq.~(\ref{eq:vendor_state}) with corresponding dimension and party number. But the idea works for general scenarios and Algorithm~\ref{alg1} can be adapted for broader classes of vendor and target states.

Once the first round is completed, all states that were saved, i.e., the states with a successful OSBP that were sufficiently close to the target state (with the information at the time), are checked to be sufficiently close with the final distribution. If not, one may try to transform them into the target via unitary transformations and an OSBP. The eventual failure branch copies can be used for estimation.

For the first rounds this protocol will do almost the same as the naive protocol. Since the distribution $p(U)$ is broad, the POVM will be close to $\{\mathds{1},0\}$, resulting most likely in the success branch (not altering the state). But as the state will be far from the target, it will be used for estimation. After many rounds the distribution $p(U)$ will be very peaked, all copies in the successful branch are now close to the target state and only the failure branches are used for estimation, resulting in a protocol similar to the one presented above.

\section{Conclusion}\label{sec:conclusion}

Potential future implementations of large-scale quantum networks, such as the envisioned `quantum internet'~\cite{Kimble2008,WehnerElkoussHanson2018,Cacciapuoti2019}, can be expected to merge idealized protocols for quantum communication and information processing with real-world conditions arising from noise and technological constraints which will necessitate the development of pragmatic and flexible solutions. Here, we have considered a quantum communication protocol that will be relevant in this context: the distribution and conversion of high-dimensional entangled states among multiple parties using noisy channels. Procedures for quantum state-conversion employed under these circumstances will involve probabilistic transformations via SLOCC. As we have shown, such scenarios are naturally amenable to enhancements via embedded channel estimation routines.

While the ideas presented here apply quite generally, we have illustrated the improvements one may expect in such metrology-assisted entanglement distribution by focusing on a particular example: probabilistic conversion of finitely many copies of $N$-qutrit states distributed via a local dephasing channels to GHZ target states. For this example, we have shown that already practically easily implementable Bayesian estimation strategies based on local measurements of individual copies can lead to excellent performance comparable to optimal distillation procedures. The latter, however, might involve significantly more complex and technologically challenging joint operations on multiple copies. 

We believe that metrology-assisted entanglement distribution has the potential to become a relevant factor in future quantum networks. We thus expect the study and optimization of such protocols for different target states and noisy models, as well as their comparison to relevant alternative protocols (e.g., based on distillation) to open up a broad range of interesting questions beyond this initial pilot study.


\begin{acknowledgments}
We thank Cornelia Spee for feedback on the manuscript. 
S.M. acknowledges support from the Austrian Science Fund (FWF) through projects Y879-N27 (START) and P 31339-N27 (Stand-Alone). 
N.F. acknowledges support from the Austrian Science Fund (FWF) through the projects Y879-N27 (START) and P 31339-N27 (Stand-Alone), and the joint Czech-Austrian project MultiQUEST (I 3053-N27 and GF17-33780L). 
M.S. ackowledges support from Spanish MINECO reference FIS2016-80681-P (with the support of AEI/FEDER,EU), the Catalan Government for the project QuantumCAT 001-P-001644 (RIS3CAT comunitats) co-financed by the European Regional Development Fund (FEDER), and 
the Generalitat de Catalunya, project CIRIT 2017-SGR-1127.  
\end{acknowledgments}


\bibliographystyle{apsrev4-1fixed_with_article_titles_full_names_new}
\bibliography{Master_Bib_File}


\appendix

\section{Details on single measurement}\label{app: single measurement}

We consider an initial measurement where each of the $N$ parties measures in the basis $\{\ket{\pm}_i=\tfrac{1}{\sqrt{2}}\bigl(\ket{0}_{i}\pm\ket{1}_{i}\bigr)\}$ for $i=1,2,\dots,N$ of the two-dimensional subspace relevant for the copies in the failure branch. Since the setup is symmetric with respect to the exchange of any of the $N$ parties, only the parity of the overall measurement matters, i.e., if one obtains an even or odd number of "$-$" outcomes. To see this note that the probability to get an outcome "$n$" corresponding to a possible operator $E_n$ is
\begin{align}
    &\tr(\proj{\Phi_{\nr\text{f}}(\theta)}E_n)=\bra{\Phi_{\nr\text{f}}(\theta)}E_n\ket{\Phi_{\nr\text{f}}(\theta)}\nonumber\\[1mm]
    &= a^2\bra{0}^{\otimes N}E_n\ket{0}^{\otimes N}
    +(1-a^2)\bra{1}^{\otimes N}E_n\ket{1}^{\otimes N}\notag\\[1mm]
    &\ \ \ \ +a\sqrt{1-a^2}\,\mathrm{e}^{-i\theta N}\bra{1}^{\otimes N}\!\!E_n\ket{0}^{\otimes N}
    \\[1mm]
    &\ \ \ \ + a\sqrt{1-a^2}\,\mathrm{e}^{i\theta N}\bra{0}^{\otimes N}\!\!E_n\ket{1}^{\otimes N},\notag
\end{align}
while $E_n$ is one of $2^N$ products of projectors $\proj{\pm}$, for instance $E_{+-+--}=\proj{+}\otimes\proj{-}\otimes\proj{+}\otimes\proj{-}\otimes\proj{-}$.
But only the parity of the number of "$-$" projector matters. Out of all $2^N$ combinations of $\proj{+}$ and $\proj{-}$ projectors for $N$ qubits, half have an even number of "$-$" projectors, so we have
\begin{align}
    p(\text{even}/\text{odd}|\theta)=\tfrac{1}{2}\pm a\sqrt{1-a^2}\cos(N\theta)
\end{align}

Given a prior $p(\theta)$, we then assign a probability
\begin{align}
    p(\text{even}/\text{odd})&=\int\limits_0^{2\pi}p(\theta)\,p(\text{even}/\text{odd}|\theta)d\theta\\[-2mm]
    &=\tfrac{1}{2}\pm a\sqrt{1-a^2}\int\limits_0^{2\pi}p(\theta)\cos(N\theta)d\theta
    \nonumber
\end{align}
and for a flat prior $p(\theta)=\frac{1}{2\pi}$ this becomes $p(\text{even}/\text{odd})=\frac{1}{2}$. 
We now use Bayes' law to obtain a posterior distribution. For a flat prior this becomes
\begin{align}
    p(\theta|\text{even}/\text{odd})=\frac{1}{2\pi}[1\pm 2a\sqrt{1-a^2}\cos(N\theta)]. 
\end{align}
Next, we nominate a suitable estimator for $N\theta$.
As an estimator we use
\begin{align}
    \hat{\theta}_N\suptiny{0}{0}{(m)}=\arg(\int\limits_0^{2\pi}p(\theta|m)\,\mathrm{e}^{iN\theta}\,d\theta),
    \label{eq:estimator}
\end{align}
and for a flat prior we have
\begin{align}
    \hat{\theta}_N^{(\text{even/odd})}=\arg(\mathrm{e}^{\frac{1\mp1}{2}i\pi}),
    \label{eq:max likelihood}
\end{align}
and so $\hat{\theta}_N^{(\text{even})}=0$ and $\hat{\theta}_N^{(\text{odd})}=\pi$.
We are then interested in evaluating the trace distance between the updated and corrected success-branch state
\begin{align}
    \rho_{\text{corr}}\suptiny{0}{0}{(m)}=\int\limits_0^{2\pi}p(\theta|m)\proj{\psi_{\text{s}}(\theta-\tfrac{\hat{\theta}_N\suptiny{0}{0}{(m)}}{N})},
\end{align}
with
\begin{align}
    \ket{\psi_{\text{s}}(\theta)}=\frac{1}{\sqrt{3}}\sum\limits_{l=0}^2\mathrm{e}^{ilN\theta}\ket{l}^{\otimes N}
\end{align}
and the desired target state $\ket{\psi_{\text{s}}(0)}$. 
The trace distance is
\begin{align}
    D(\rho_{\text{corr}}\suptiny{0}{0}{(m)},\rho_{\text{target}})=\frac{1}{2}\tr\bigg(\sqrt{(\rho_{\text{corr}}\suptiny{0}{0}{(m)}-\rho_{\text{target}})^2}\bigg),
\end{align}
and we can calculate
\begin{align}
    \rho_{\text{corr}}\suptiny{0}{0}{(m)}-\rho_{\text{target}}=\frac{1}{3}\sum\limits_{j\ne l}c_{jl}\suptiny{0}{0}{(m)}(\ketbra{j}{l})^{\otimes N},
\end{align}
with
\begin{align}
    c_{jl}\suptiny{0}{0}{(m)}=\int\limits_0^{2\pi}p(\theta|m)\,\mathrm{e}^{i(N\theta-\hat{\theta}_N\suptiny{0}{0}{(m)})(j-l)}d\theta -1.
\end{align}
For $\hat{\theta}_N\suptiny{0}{0}{(m)}=0$ and $\hat{\theta}_N\suptiny{0}{0}{(m)}=\pi$ the coefficients $c_{jl}\suptiny{0}{0}{(m)}$ are real and therefore
\begin{align}
    D(\rho_{\text{corr}}\suptiny{0}{0}{(m)},\rho_{\text{target}})=\frac{1}{6}(|c_{02}|+\sqrt{8c_{01}^2+c_{02}^2}).
\end{align}
For the flat prior we find 
\begin{subequations}
\begin{align}
    c_{0,1}^{(\text{even}/\text{odd})}  &=\,c_{1,2}^{(\text{even}/\text{odd})}\,=\,a\sqrt{1-a^2}-1,\\[1mm]
    c_{0,2}^{(\text{even}/\text{odd})}  &=\,-1,
\end{align} 
\end{subequations}
and with this the trace distance becomes
\begin{small}
\begin{align}
    D(\rho_{\text{corr}}\suptiny{0}{0}{(m)},\rho_{\text{target}})=
    \tfrac{1}{6}\Bigl(1\!+\!\sqrt{9\hspace*{-0.5pt}+\hspace*{-0.5pt}8a^2(1\hspace*{-0.5pt}-\hspace*{-0.5pt} a^2)\hspace*{-0.5pt}-\hspace*{-0.5pt}16a\sqrt{1\hspace*{-0.5pt}-\hspace*{-0.5pt} a^2}}\,\Bigr),
\end{align}
\end{small}\noindent and since this is independent of the outcome it already matches the average trace distance $\bar{D}(\rho_{\text{corr}},\rho_{\text{target}})=\sum\limits_m p(m) D(\rho_{\text{corr}}\suptiny{0}{0}{(m)},\rho_{\text{target}})$.


\section{Multiple copies without updating}
\label{app: multiple copies}

Now let us see how measurements on more than one copy, in particular on $k$ copies, can improve this value by considering a simple scenario where the measurement directions are not updated. This means, each copy is locally measured in the basis $\{\ket{\pm}_{i}\}$ by every party, resulting in $k$ individual even/odd measurement outcomes. Since there is no updating in-between, the particular order of these outcomes is irrelevant and only the overall number of even/odd outcomes matter. We can thus view the entire procedure as one $(k+1)$-outcome measurement with outcomes $m=0,1,2,\dots,k$, i.e., the number of "even" outcomes. 
We hence assign the probabilities
\begin{align}
    p(m|\theta)=p(\text{even}|\theta)^m p(\text{odd}|\theta)^{k-m}\begin{pmatrix}k\\m\end{pmatrix}.
    \label{eq:likelihood_no_updating}
\end{align}
The corresponding unconditional probabilities are
\begin{align}
    p(m)=&\frac{1}{2^k}\begin{pmatrix}k\\m\end{pmatrix}\sum\limits_{i=0}^{m}\sum\limits_{j=0}^{k-m}\begin{pmatrix}m\\i\end{pmatrix}\begin{pmatrix}k-m\\j\end{pmatrix}(-1)^j\notag\\
    &\ \times\,(2a\sqrt{1-a^2})^{i+j}\,\int\limits_0^{2\pi}p(\theta)\,\cos^{i+j}(N\theta)d\theta.
    \label{eq:prob_outcome_no_updating}
\end{align}
For a flat prior we calculate
\begin{align}
    p(m)=&\frac{1}{2^k}\begin{pmatrix}k\\m\end{pmatrix}\sum\limits_{i=0}^{m}\sum\limits_{j=0}^{k-m}\begin{pmatrix}m\\i\end{pmatrix}\begin{pmatrix}k-m\\j\end{pmatrix}(-1)^j\notag\\
    &\ \times\,(2a\sqrt{1-a^2})^{i+j}\,\frac{(-1)^i+(-1)^j}{2^{i+j+1}}\begin{pmatrix}i+j\\\frac{i+j}{2}\end{pmatrix}.
    \label{eq:prob_outcome_no_updating_flatprior}
\end{align}
The posterior given outcome $m$ is then
\begin{align}
    p(\theta|m)=&\frac{1}{2^kp(m)}\begin{pmatrix}k\\m\end{pmatrix}\sum\limits_{i=0}^{m}\sum\limits_{j=0}^{k-m}\begin{pmatrix}m\\i\end{pmatrix}\begin{pmatrix}k-m\\j\end{pmatrix}\notag\\
    &\times\,(-1)^j\,(2a\sqrt{1-a^2})^{i+j}\,p(\theta)\,\cos^{i+j}(N\theta).
    \label{eq:posterior_no_updating}
\end{align}
For the estimator we have to calculate
\begin{align}
    \bar{z}_{N,k}\suptiny{0}{0}{(m)}=\int\limits_0^{2\pi}p(\theta|m)\mathrm{e}^{iN\theta}d\theta.
    \label{eq:estimator_no_updating}
\end{align}
\begin{figure}[ht!]
     \includegraphics[width=0.95\columnwidth]{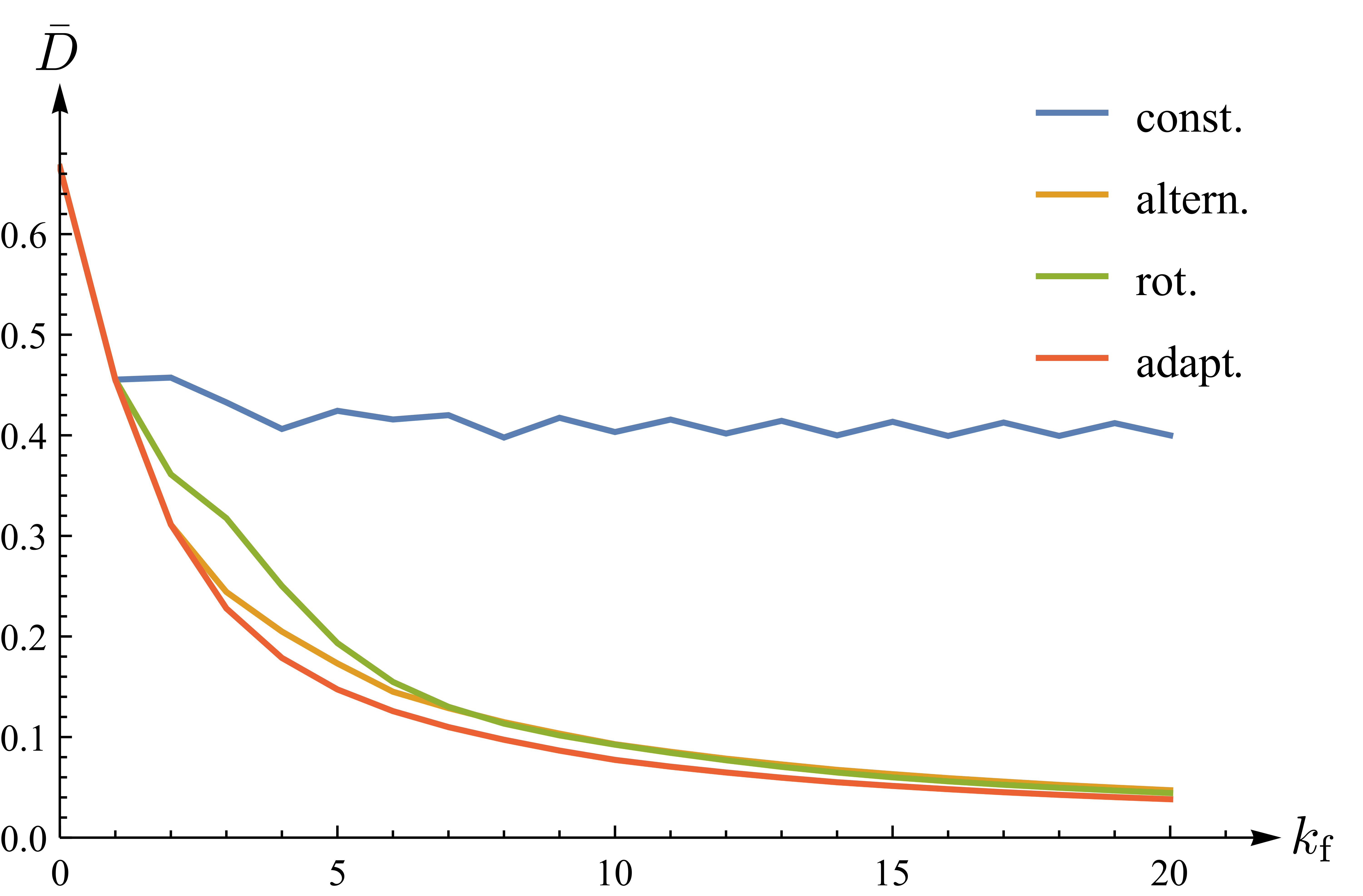}
        \caption{
        Average distance to target state for different ways of adapting the measurement direction. 
        The curves show the average trace distance between the target state and the successfully converted copies for different measurement-updating strategies for the exemplary metrology-assisted entanglement distribution protocol discussed in Sec.~\ref{sec:dephasing_noise} for fixed value of $a=\tfrac{1}{\sqrt{2}}$ as a function of the number $k_{\text{f}}$ of copies in the failure branch. The blue curve represents the strategy with constant measurement directions discussed in Appendix~\ref{app: multiple copies}, for the yellow curve measurements are alternated between $\sigma_x$ and $\sigma_y$, for green the measurement directions are rotated by a constant angle $\pi/8$ after each measurement, and the red curve corresponds to an adaptive strategy updating the measurement direction using the maximum likelihood estimator from Eq.~(\ref{eq:max likelihood}) of the previous round.
        }
    \label{fig:distance3}
\end{figure}
For the flat prior the estimator becomes the argument of
\begin{align}
    \bar{z}_{N,k}\suptiny{0}{0}{(m)}=&\frac{1}{2^kp(m)}\begin{pmatrix}k\\m\end{pmatrix}\sum\limits_{i=0}^{m}\sum\limits_{j=0}^{k-m}\begin{pmatrix}m\\i\end{pmatrix}\begin{pmatrix}k-m\\j\end{pmatrix}(-1)^j\notag\\
    &\ \times\,(2a\sqrt{1-a^2})^{i+j}\,\frac{(-1)^i+(-1)^j}{2^{i+j+1}}\begin{pmatrix}i+j\\\frac{i+j}{2}\end{pmatrix}.
    \label{eq:posterior_without_updating_flatprior}
\end{align}
We can distinguish two cases, as long as $m<k/2$ we have $ \bar{z}_{N,k}\suptiny{0}{0}{(m)}<0$ and therefore $\arg( \bar{z}_{N,k}\suptiny{0}{0}{(m)})=\pi=\hat{\theta}_{N,m}\suptiny{0}{0}{(m)}$. For $m\ge k/2$ we have $ \bar{z}_{N,k}\suptiny{0}{0}{(m)}\ge0$ and therefore $\arg( \bar{z}_{N,k}\suptiny{0}{0}{(m)})=0=\hat{\theta}_{N,m}\suptiny{0}{0}{(m)}$. What remains is calculating the distance to our target state, and the result is illustrated in Fig.~\ref{fig:distance3}.

In the same figure we numerically investigate the performance of different measurement-updating strategies. We notice that, perhaps unsurprisingly, the strategy with constant measurement direction performs worst of all considered strategies, whereas strategies with changing measurement directions perform considerably better.
More specifically, we compare an alternating strategy that switches between two mutually unbiased measurements, and a rotating strategy, where the measurement direction is rotated by a fixed angle, to the adaptive strategy. In the latter strategy, which is also considered in the main text, the measurement direction is updated based on the outcome of the previous measurement rounds and chosen perpendicular (in Bloch representation) to the estimator for the angle $\theta$ from the previous round.
While the adaptive strategy performs best, as expected, also the strategies that change the measurement direction according to some pre-established rule perform comparably well. This gives viable alternatives that are possibly easier to implement, as they do not depend on previous outcomes.


\section{Estimation with received copies}
\label{app: good copy estimation}

\begin{figure}[ht!]
     \includegraphics[width=0.95\columnwidth]{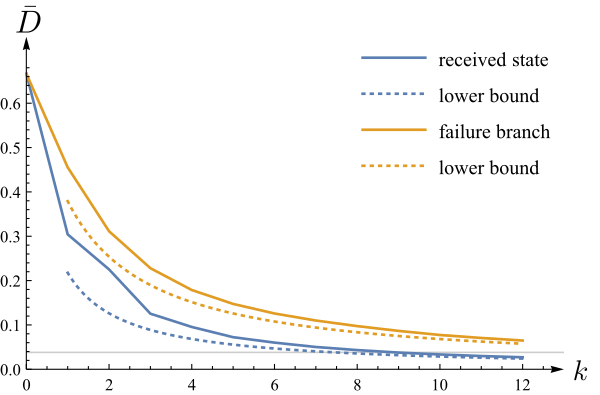}
        \caption{
        Average distance to target state for estimation with received states.
        The curve shows the average trace distance between the target state and the successfully converted copies as a function of the number $k$ of copies used. The blue line shows the 'naive' protocol discussed in Sec.~\ref{sec:comparison}, where the received states $\ket{\Psi(\alpha,\beta)}$ from Eq.~(\ref{eq:vendor_state2}) are used directly for the estimation.
        The yellow curve shows the estimation performed on copies of the failure branch described by Eq.~(\ref{eq:fail_state}).
        The gray line shows the trace distance of $\epsilon\approx0.038$ after measuring 20 copies of the failure branch and we see that 9 copies of the received state are necessary to achieve the same distance.
        The dotted line represents the lower bound calculated in Appendix~\ref{app: lower bound good copy estimation} based on the Bayesian Cram\'er-Rao bound for the respective probe state.
        The parameters used are $\alpha=\arccos(2/\sqrt{15})$ and $\beta=\pi/4$.
        }
    \label{fig:distance4}
\end{figure}

In this appendix we briefly explain how to estimate the phase $\theta$ using the state $\ket{\Psi(\alpha,\beta)}$ given in Eq.~(\ref{eq:vendor_state2}) along with a Bayesian estimation strategy. The strategy is essentially the same as that employed before, adapted to the qutrit probe states.
First, we choose an appropriate three-outcome projective measurement in a basis that is mutually unbiased with respect to the eigenbasis of the generating Hamiltonian of the transformation $U(\theta)$ from Eq.~(\ref{eq:unitaries}), e.g. the discrete Fourier transform of the standard basis
\begin{subequations}
\begin{align}
    \ket{\xi_0}&=\tfrac{1}{\sqrt{3}}(\ket{0}+\ket{1}+\ket{2}),\\[1mm]
    \ket{\xi_1}&=\tfrac{1}{\sqrt{3}}(\ket{0}+\omega\ket{1}+\omega^2\ket{2}),\\[1mm]
    \ket{\xi_2}&=\tfrac{1}{\sqrt{3}}(\ket{0}+\omega^2\ket{1}+\omega\ket{2}),
\end{align}
\end{subequations}
where $\omega=\mathrm{e}^{2\pi i/3}$. Each party performs a measurement in this basis. Let $E_n$ denote the $3^N$ possible products of $N$ projectors $\ketbra{\xi_j}{\xi_j}$. 
We then calculate
\begin{align}
    &\tr(\proj{\Psi(\alpha,\beta)}E_n)=\brakket{\Psi(\alpha,\beta)}{E_n}{\Psi(\alpha,\beta)}\notag\\[1mm]
    &= \tfrac{1}{3^N}\big[1+2\sin^2\alpha\cos\beta\sin\beta \cos(\tfrac{2\pi}{3}[n_1+2n_2]-\theta N)\notag\\[1mm]
    &\ \ +2\sin\alpha\cos\alpha\cos\beta \cos(\tfrac{2\pi}{3}[2n_1+n_2]-2\theta N)\notag\\[1mm]
    &\ \ +2\sin\alpha\sin\beta\cos\alpha \cos(\tfrac{2\pi}{3}[n_1-n_2]-\theta N)\big], 
\end{align}
where $n_j$ denotes the number of projectors $\ketbra{\xi_j}{\xi_j}$ in $E_n$. 
The probabilities for the different outcomes of the measurement hence only depend on the respective number of projectors constituting the measurement, which also follows from the symmetry with respect to the exchange of the $N$ parties. 
Moreover, the probabilities of the outcomes only depend on $n_1+2n_2$, $2n_1+n_2$ and $n_1-n_2$ modulo $3$. We therefore only need to distinguish between three possible measurement outcomes $(0,0,0)$, $(1,2,1)$, and $(2,1,2)$, which we can rewrite as $m^l=(l,2l,l)$.
It is clear that there are $3^{(N-1)}$ sequences resulting in a given outcome, and hence it follows that
\begin{align}
    p(m^l|\theta)
    &=\, \tfrac{1}{3}\big[1+2\sin^2\alpha\cos\beta\sin\beta
    \cos(\tfrac{2\pi}{3}l-\theta N)\notag\\[1mm]
    &\ +2\sin\alpha\cos\alpha\cos\beta \cos(\tfrac{4\pi}{3}l-2\theta N)\notag\\[1mm]
    &\ +2\sin\alpha\sin\beta\cos\alpha \cos(\tfrac{2\pi}{3}l-\theta N)\big].
\end{align}
Here, we again want to mention that we actually estimate $\theta N$ and that the estimation precision is therefore independent of the number of parties.
By Bayes' law we can now calculate $p(\theta|m)$ and use the estimator from Eq.~(\ref{eq:estimator}).
In each subsequent measurement round we adapt the measurement direction based on the outcome of the previous round, that is we rotate our measurement basis by $U(\hat{\theta}+\pi/3)$.
In Fig.~\ref{fig:distance4} we compare the resulting average trace distance for this strategy to that obtained from an estimation strategy based only on the failure-branch states.


\section{A lower bound for the number of copies for any estimation protocol}
\label{app: lower bound good copy estimation}

In this appendix we derive a lower bound for the number of copies needed for any estimation to achieve a given trace distance. 
Since we assume that no prior information about the phase is available initially, we rely on Bayesian estimation with a flat prior distribution. 
After $k$ measurement rounds the posterior distribution of the parameter $\theta$ is $p(\theta|m_1,\dots,m_{k})=p(\theta|\mathbf{m})$ and the corrected state becomes
\begin{align}
    \rho_{\text{corr}}\suptiny{0}{0}{(\mathbf{m})}=\int\limits_0^{2\pi}p(\theta|\mathbf{m})\proj{\psi_{\text{s}}(\theta-\tfrac{\hat{\theta}_N\suptiny{0}{0}{(\mathbf{m})}}{N})}d\theta,
\end{align}
for some estimator $\hat{\theta}_N\suptiny{0}{0}{(\mathbf{m})}$.
Now we can bound the average trace distance from below using its convexity, i.e., using the notation $\rho_{\text{target}}=\proj{\psi_{\text{s}}(0)}$ and $\rho_{\text{s}}(\theta)=\proj{\psi_{\text{s}}(\theta)}$ we have
\begin{align}
    &\bar{D}(\rho_{\text{corr}}\suptiny{0}{0}{(\mathbf{m})},\rho_{\text{target}})
    =\sum\limits_\mathbf{m} p(\mathbf{m}) D(\rho_{\text{corr}}\suptiny{0}{0}{(\mathbf{m})},\rho_{\text{target}})\label{eq:inequality1}\\
    &\ \ge D(\sum\limits_\mathbf{m}p(\mathbf{m})\rho_{\text{corr}}\suptiny{0}{0}{(\mathbf{m})},\rho_{\text{target}})\nonumber\\
    &\ \ =D\bigl(\,
    \int\limits_0^{2\pi}\sum\limits_\mathbf{m}p(\mathbf{m})p(\theta|\mathbf{m})\rho_{\text{s}}
    (\theta-\tfrac{\hat{\theta}_N\suptiny{0}{0}{(\mathbf{m})}}{N})d\theta,\,\rho_{\text{target}}\,\bigr).\nonumber
\end{align}
To continue the calculation and to actually compute a useful lower bound we need to make an approximation. We replace the average posterior distribution by a Gaussian distribution with the same mean $\mu$ and variance $\sigma^2$.
Since the trace distance $D(\ket{\psi_{\text{s}}(\theta)},\ket{\psi_{\text{s}}(0)}$ grows sublinearly with $\theta$, we expect that the difference in the higher-order moments of the distributions will not influence the result significantly. Additionally, after some number of measurements the estimation is sufficiently precise such that the numerical support of the distribution is within the interval $[0,2\pi)$ and we can therefore integrate over the real line. 
With this we finally conclude
\begin{align}
    &\bar{D}(\rho_{\text{corr}}\suptiny{0}{0}{(\mathbf{m})},\rho_{\text{target}})\ge D(\sum\limits_\mathbf{m}p(\mathbf{m})\rho_{\text{corr}}\suptiny{0}{0}{(\mathbf{m})},\rho_{\text{target}})\nonumber\\
    &\approx D\bigl(
    \int\limits_{-\infty}^{\infty}\!\!\frac{1}{\sqrt{2\pi}\sigma}\mathrm{e}^{-\frac{(\theta-\mu)^2}{2\sigma^2}}
    \rho_{\text{s}}
    (\theta-\tfrac{\hat{\theta}_N\suptiny{0}{0}{(\mathbf{m})}}{N})d\theta,\,\rho_{\text{target}}\,\bigr)\nonumber\\
    &\ge D\bigl(
    \int\limits_{-\infty}^{\infty}\!\!\frac{1}{\sqrt{2\pi}\sigma}\mathrm{e}^{-\frac{(\theta-\hat{\theta}_N\suptiny{0}{0}{(\mathbf{m})}/N)^2}{2\sigma^2}} \rho_{\text{s}}
    (\theta-\tfrac{\hat{\theta}_N\suptiny{0}{0}{(\mathbf{m})}}{N})d\theta,\,\rho_{\text{target}}\,\bigr)\nonumber\\
    &=\tfrac{\mathrm{e}^{-2N^2\sigma^2}}{6}
    \bigl(\mathrm{e}^{\tfrac{N^2\sigma^2}{2}}\!-\! 1\bigr)
    \Bigl[1\!+\!\mathrm{e}^{\tfrac{N^{2}\sigma^{2}}{2}}\!+\!\mathrm{e}^{N^2\sigma^2}+\mathrm{e}^{\tfrac{3N^2\sigma^2}{2}}\notag\\[1mm]
    &\ \ +\sqrt{8\mathrm{e}^{3N^2\sigma^2}+(1+\mathrm{e}^{N^2\sigma^2/2})^2(1+\mathrm{e}^{N^2\sigma^2})^2}\,\Bigr],
    \label{eq:inequality5}
\end{align}
where we have used the fact that the trace distance is minimal if the mean of the Gaussian distribution equals the value of the estimator, i.e., using the mean of the posterior as estimator. 
The average variance of any estimator is bounded from below by the van Trees inequality $V[\hat{\theta}(\mathbf{m})]\ge1/(I[p(\theta)]+\bar{I}[p(\mathbf{m}|\theta)])$, where $I[p(\theta)]$ is the Fisher information (FI) of the prior, and $\bar{I}[p(\mathbf{m}|\theta)])$ is the average FI of the likelihood over all values of $\theta$~\cite{GillLevit1995}. Now we bound the FI from above by the quantum FI (QFI), $p(\mathbf{m}|\theta)]\ge\mathcal{I}[\ket{\psi(\theta)}]$, and notice that the QFI is independent of $\theta$ for a unitary evolution and hence equals its average~\cite{BraunsteinCaves1994,DemkowiczDobrzanskiJarzynaKolodynski2015}. One can thus get a Bayesian analogue of the quantum Cram\'er-Rao bound for the mean of the posterior, bounding the average posterior variance by $\bar{V}[p(\theta|\mathbf{m})]\ge1/(I[p(\theta)]+\mathcal{I}[\ket{\psi(\theta)}])$ (see, e.g.,~\cite{FriisOrsucciSkotiniotisSekatskiDunjkoBriegelDuer2017,MorelliUsuiAgudeloFriis2021}). For the state in Eq.~(\ref{eq:vendor_state2}) the FI is
\begin{align}
    \mathcal{I}\bigl[\ket{\psi(\theta)}\bigr]
    &= \,4 N^2 [\sin^2\alpha\sin^2\beta+4\cos^2\alpha\notag\\[1mm]
    &\ \ \ \ -(\sin^2\alpha\sin^2\beta+2\cos^2\alpha)^2]
    \label{eq:Fischer_information_vendor_state}
\end{align}
and for a flat prior the FI vanishes. 
Since the expression in the last line of~(\ref{eq:inequality5}) increases monotonically in $\sigma^2$ we can further bound the average trace distance from below by inserting $\sigma^2=1/(k\ \mathcal{I}\bigl[\ket{\psi(\theta)}\bigr])$. This leaves us with an expression that is decreasing monotonically with the number $k$ of copies used for estimation but which is independent of the number of parties $N$. With this we can determine a lower bound on the number of copies needed to reach a given trace distance.

\end{document}